# Potentially Significant Source of Error in Magnetic Paleolatitude Determinations

J. Marvin Herndon
Transdyne Corporation
San Diego, CA 92131 USA
mherndon@san.rr.com

**Abstract:** The discovery of close-to-star gas-giant exo-planets lends support to the idea of Earth's origin as a Jupiter-like gas giant and to the consequences of its compression, including whole-Earth decompression dynamics that gives rise, without requiring mantle convection, to the myriad measurements and observations whose descriptions are attributed to plate tectonics. I show here that paleolatitude determinations, used extensively in Pangaea-like reconstructions and in paleoclimate considerations, may be subject to potentially significant errors if rock-magnetization was acquired at Earth-radii less than present.

Deciphering the record of Earth's ancient magnetic field, imprinted in the minerals of rocks during cooling and/or chemical alteration and/or deposition, has wide-ranging applicability and importance. For decades, estimates of rock-sample latitude at the time of acquired magnetization, called paleolatitude, have been deduced from measurements the extant magnetic inclination. Paleolatitude determinations provide the principal basis for Pangaea-like supercontinent reconstructions and are used extensively in paleoclimate considerations. Consequently, great efforts are made to ascertain and eliminate paleolatitude bias-causing factors [1, 2]. Here I show in a general framework a potentially significant bias in paleolatitude estimates that might arise from determinations made on rock-samples that became magnetized at Earth-radii less than the present value, circumstances that new investigations reveal to be quite possible [3].

The idea that Earth's radius may have been smaller in the past is not new. In 1933, Hilgenberg [4] envisioned one continent without ocean basins on a globe smaller than Earth's present diameter that subsequently expanded in a process that fragmented and separated continental masses and formed interstitial ocean basins. Hilgenberg's concept provided the basis for "Earth expansion theory" [5]. But Earth expansion theory as formulated is unable to explain the reason for Earth's initially smaller size or to provide a source for the vast energy required for expansion. Furthermore, the idea that Earth expansion had occurred solely within the past 170 million years, the age of the oldest seafloor, is at odds with geological evidence. Moreover, Earth expansion theory is unable to provide explanations for seafloor topography that are well-described by plate tectonics theory. But, for all of its attractive features, plate tectonics theory has underlying problems too, especially being crucially dependent upon the problematic concept of mantle convection.

I have united "plate tectonics" and "Earth expansion" into a new geodynamic theory, called *whole-Earth decompression dynamics*, that describes the consequences of our planet's early formation as a Jupiter-like gas giant [3, 6, 7] and gives rise, without requiring mantle convection, to the myriad measurements and observations whose descriptions are attributed to plate tectonics.

Envision pre-Hadean Earth, compressed to about 64% of present radius by about 300 Earth masses of primordial gases and ices. At some point, after being stripped of its massive volatile envelope, presumably by the Sun's super-intense T-Tauri solar winds, internal pressures would build eventually cracking the rigid crust. Powered by the stored energy of protoplanetary compression, Earth's progressive decompression is manifest at the surface by the formation of cracks: *primary* decompression cracks with underlying heat sources capable of extruding basalt, and *secondary* decompression cracks without heat sources that serve as ultimate repositories for basalt extruded from primary decompression cracks. Mid-ocean ridges and submarine trenches, respectively, are examples of these. Secondary decompression cracks serve to increase surface area in response to decompression-driven volume expansion. Basalt extruded at mid-ocean ridges becomes seafloor, spreading and eventually subducting, i.e., falling into secondary decompression cracks, seismically imaged as "down-plunging slabs", but without engaging in the process of mantle convection.

The supposition that Earth's spin leads to magnetic poles being near geographic poles irrespective of Earth's past smaller radius, except during periods of reversals, is reasonable, but may lead the (false) conclusion that rock-magnetization acquired at one radius value would have the same direction at a later increased radius, if no "continental drift" occurred.

Figure 1 shows a hypothetical 4000km "ancient" continent cross-section (arc *ACE*) at a radius of 64% of Earth's present radius, *OC,* and the same "present" continent cross-section (arc *GHI*) at present Earth radius, *OH*. Consider the line *OR* as a fixed reference, not necessarily a pole, but relatable to a pole. In Figure 1, clearly no "continental drift" has occurred, as the reference line *OR* bisects both the ancient and present continent. Significantly, the ancient continent subtends an angle, <*AOE*, that is considerable greater than the angle the present continent subtends, <*GOI*: 56.3degrees versus 36.0 degrees.

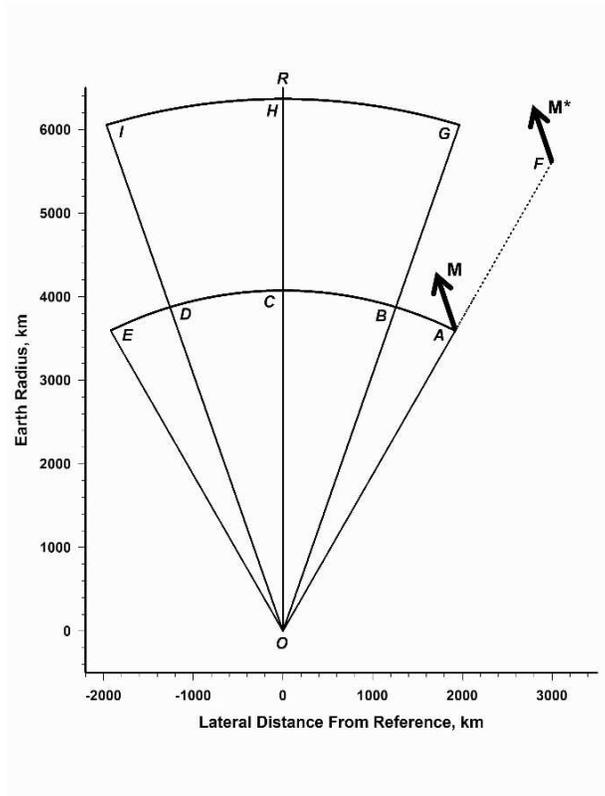

**Figure 1**. Cross-section of a hypothetical continent, 4000km long, at a time when Earth's radius was 64% of the present radius and at present. Details are described in the text.

Consider a magnetization direction imprinted in the magnetic minerals of a rock at an arbitrary point along the ancient continent cross-section, (arc *ACE*). For clarity, here assume that the magnetized rock-unit is located at the continent's edge, point *A*, and has magnetization direction indicated by **M**. Because of the decompression-driven increase in planetary radius and concomitant increase in ocean-floor surface area, the direction of the ancient magnetization, **M**, when observed later at point *G*, will appear to have been acquired at a different paleolatitude. To illustrate, imagine moving the rock-unit with its acquired magnetization to point *G* in a two-step process. Imagine first moving the rock-unit the distance *OG* along the radial extension *OF*; note that its magnetization, **M***, at point *F* is parallel to **M**. Clearly, the second movement of the rock-unit to bring it to point *G* will involve closing <*FOG*, thus rotating the apparent direction of **M** by <*FOG*; in this example by 10.1 degrees.

As shown in the above hypothetical example, significant potential bias in paleolatitude determinations my arise as a consequence of magnetization having been acquired at Earth radii less than present value. In the case of no "continental drift", as inferred from Figure 1, the magnitude of the bias, though, should diminish as sampling approaches mid-continent. A second potential source of bias, one more difficult to quantify, may arise from internal adjustments related to changes in curvature. In the example, the present cord length, *GI*, is 93km longer than the ancient cord length, *BD*; concomitantly, the maximum rise above the cord at mid-continent is 170km less in the present than in the ancient.

It is not the purpose here to debate the question of whether Earth had a shorter radius in the past, but rather to point to a potential source of error in paleolatitude determinations that, once recognized, may lead to important discoveries. Good science demands considering all potential sources of bias.

## Acknowledgement

I thank Professor Lynn Margulis for encouragement and advice and I thank author-photographer Reg Morrison for inspiring the present work.